\documentclass[useAMS,usenatbib,usegraphicx]{mn2e}

\usepackage{color, aas_macros, pifont}
\title[Continuous MHD oscillation spectrum of a magnetically confined mountain]{Continuous frequency spectrum of the global hydromagnetic
  oscillations of a magnetically confined mountain on an accreting
  neutron star}

\def\eref#1{equation (\ref{eq:#1})}
\def\eeref#1{(\ref{eq:#1})}
\def\fref#1{Fig. \ref{fig:#1}}
\def\sref#1{section \ref{sec:#1}}

\def\bcdot{\bmath{\cdot}}
\def\change#1{{#1}}
\def\matrix#1{\mathbfss{#1}}

\def\ri{\mathrm{i}}
\def\flone{\ding{192}}
\def\fltwo{\ding{193}}
\def\flthree{\ding{194}}
\def\flfour{\ding{195}}

\author[M. Vigelius et al.]{M.~Vigelius $^1$\thanks{E-mail: mvigeliu@physics.unimelb.edu.au} and A.~Melatos $^1$ \\
 $^1$ School of Physics, University of Melbourne, Parkville, VIC
 3010, Australia}

\begin{document}

\date{Submitted to MNRAS}

\maketitle

\begin{abstract}
We compute the continuous part of the ideal-magnetohydrodynamic (ideal-MHD)
frequency spectrum of a polar mountain produced by magnetic burial on
an accreting neutron star. Applying the formalism developed by
\citet{Hellsten79}, extended to include gravity, we solve the singular eigenvalue
problem subject to line-tying boundary conditions. This spectrum
divides into an Alfv\'{e}n part and a cusp part.
The eigenfunctions are chirped and anharmonic with an exponential
envelope, and the eigenfrequencies cover the whole spectrum above a minimum
$\omega_\mathrm{low}$. For equilibria with accreted
mass $1.2 \times 10^{-6} \la M_a/M_\odot \la 1.7 \times 10^{-4}$ and
surface magnetic fields $10^{11} \la B_\ast/\mathrm{G} \la 10^{13}$,
$\omega_\mathrm{low}$ is approximately independent of $B_\ast$, and
increases with $M_a$. The results are consistent with the Alfv\'{e}n
spectrum excited in numerical simulations with the \textsc{zeus-mp}
solver. The spectrum is modified substantially by the Coriolis force
in neutron stars spinning faster than $\sim 100$ Hz. The implications
for gravitational wave searches for low-mass X-ray binaries are
considered briefly.
\end{abstract}

\bibliographystyle{mn2e}


\section{Introduction}
\label{sec:intro}
Radio and X-ray observations suggest that the
magnetic dipole moments, $\mu$, of  neutron stars in accreting
binaries decrease with accreted mass, $M_a$
\citep{Taam86, vanDenHeuvel95}. One physical mechanism capable of
reducing $\mu$ by the amount observed is
magnetic screening or burial
\citep{Bisnovatyi74,Romani90,Konar97,Zhang98,Melatos01,Choudhuri02,Payne04,Lovelace2005}.
Magnetic burial occurs when accreting plasma, flowing inside the
Alfv\'{e}n radius, is channelled onto the
magnetic poles of the neutron star. The hydrostatic pressure at the
base of the accreted column overcomes
the magnetic tension and the column spreads equatorwards, distorting the frozen-in magnetic flux
\citep{Melatos01}. Self-consistent magnetohydrodynamic (MHD) equilibria respecting the
flux-freezing, ideal-MHD constraint were computed by
\citet{Payne04}, who found that the magnetic field is compressed into an equatorial belt,
which confines the accreted mountain at the poles. A key result is
that $\mu$ drops significantly once $M_a$ exceeds
the critical mass $M_c \sim 10^{-5} M_\odot$. This characteristic value exceeds simple
estimates based on \emph{local} MHD force balance at the polar cap
, i.e. without the equatorial magnetic belt \citep{Brown98, Litwin01}.

Counterintuitively, magnetic mountain equilibria prove to be marginally stable
\citep{Payne07, Vigelius08}. An axisymmetric mountain is
susceptible to the   undular submode of the Parker instability, but
the instability is transitory, saturating after $\sim 10$ Alfv\'{e}n
times to give a nearly axisymmetric, mountain-like state, which oscillates
in a superposition of small-amplitude, global, Alfv\'{e}n and acoustic
modes \citep{Payne07, Vigelius08}. The magnetic line-tying boundary
condition at the stellar surface is crucial in stabilizing the
mountain. 

The Alfv\'{e}n and acoustic oscillations help to shape the gravitational wave  spectrum
emitted by low-mass X-ray binaries \citep{Vigelius08b}, furnishing a
new observational probe of the surface magnetic structure
of neutron stars. The oscillations may also manifest themselves as
small, Hz-to-kHz variations in the X-ray
pulse shape. Accordingly, a linear eigenmode
analysis of the ideal-MHD spectrum is required to take full advantage
of future gravitational-wave and X-ray timing
experiments. Here, we take a first step by computing the
\emph{continuous} part of the ideal-MHD 
spectrum analytically for axisymmetric magnetic mountains, closely following the
approach of \citet{Hellsten79}.

\change{Realistically, however, the detection of the
  gravitational-wave imprint from mountain oscillations will not be
  possible in the near future. While Advanced Laser Interferometer
  Gravitational Wave Observatory (Advanced LIGO) may detect the
  unperturbed mountain \citep{Watts08, Vigelius08b}, the detection of
  modulations of the main signal from mountain oscillations will have
  to await next-generation interferometers. On the other hand,
  X-ray-profile changes can currently be measured with an accuracy of
  $\sim 1$ per cent \citep{Muno02, Hartman08} and we know that the
  accretion rate changes by $\sim 10$ per cent per day. Unfortunately,
  the X-ray fluctuations will be only poorly frequency-matched to the
  mountain oscillations in general.}

Generally, an inhomogenuous MHD configuration supports
linear eigenmodes \citep{Lifschitz89,Goedbloed04}, whose frequency
spectrum divides into a discrete and a
continuous part. Discrete eigenvalues are fixed by the boundary
conditions. Continuous eigenvalues arise from singularity
in the underlying Sturm-Liouville problem, which allows the boundary
conditions to be satisfied for any eigenvalue within a range.

In earlier work, stochastically excited mountain oscillations were
investigated numerically by perturbing equilibrium stars with
different $M_a$ in the ideal-MHD solver \textsc{zeus-mp}
\citep{Hayes06}, and computing
the spectrum \citep{Payne07, Vigelius08}. However, this approach is
restricted to the subset of the full MHD spectrum resolved by
\textsc{zeus-mp} and is computationally expensive. In this article, we 
attack the problem analytically. The article is organised as
follows. In \sref{equilibrium}, we
introduce curvilinear field line coordinates to describe the
equilibrium and establish the associated
metric. We derive the linearized, ideal-MHD equations in these coordinates in
\sref{linearized}, extending the analysis by \citet{Hellsten79} to
include the gravitational field of a central point mass. The
continuous frequency spectrum and the corresponding eigenfunctions are evaluated
in \sref{spectrum}, as a function of $M_a$ and the magnetic field strength
before burial. We conclude by discussing the implications for
gravitational wave observations of accreting millisecond pulsars in
\sref{discussion}.

\section{Hydromagnetic equilibrium}
\label{sec:equilibrium}

The equilibrium structure of a magnetically confined mountain in ideal
MHD is described by the force balance equation
\begin{equation}
  \label{eq:equilibrium:force_balance}
  \nabla P - (\nabla \times
  \bmath{B})\times \bmath{B}= N \bmath{g},
\end{equation}
supplemented by $\nabla \bcdot \bmath{B}=0$ and an
equation of state $P(N)$, which we take to be isothermal: $P=c_s^2
N$. Here, P, $\bmath{B}$,
$N$, $\bmath{g}$, and $c_s$ denote the pressure, magnetic field, mass density, 
gravitational acceleration, and isothermal sound speed, respectively.

If we introduce a cylindrical coordinate system 
$(r,\varphi,z)$ and assume axisymmetry, we can write
\begin{equation}
  \label{eq:equilibrium:coordinates}
  \bmath{B}=\nabla \varphi \times \nabla \Psi,
\end{equation}
where $\Psi$ is the magnetic flux, measured in G cm$^2$. \citet{Payne04} computed unique,
self-consistent, MHD equilibria by solving
\eeref{equilibrium:force_balance} in spherical coordinates subject to
the flux-freezing constraint of ideal MHD. In this article,
we convert these solutions to cylindrical coordinates
before constructing the associated field line coordinates.

Throughout this article, we work in cgs-like units, such that
$\mu_0=1$. Furthermore, we normalize the isothermal sound speed and
the gravitational constant, viz. $c_s=G=1$.

\subsection{Field line coordinates}
A magnetic mountain is created by
continuously deforming a dipole magnetic field during the accretion
process. The flux surfaces are closed at all times, a topology that
is preserved even for a uniform $B_\varphi\ne0$. We can therefore
introduce orthogonal curvilinear coordinates $(\Psi, \eta, \varphi)$,
called field line coordinates, which ``follow'' the shape of the flux
surface. In these coordinates, $\eta$ measures arc length along a magnetic field
line, normalized to the domain $0 \le \eta \le 1$, where $\eta=0$
corresponds to the footpoint at the surface, and $\eta=1$ corresponds to the
outer radial boundary or, when the field line is closed, to the
surface. $\Psi$ is the flux function in \eeref{equilibrium:coordinates} 
and $\varphi$ is the usual azimuthal angle in cylindrical coordinates. In the
field line coordinates, the metric becomes \citep{Hellsten79}
\begin{equation}
  \mathrm{d}s^2 = \frac{1}{r^2 B^2} \mathrm{d}\Psi^2+J^2 B^2
  \mathrm{d}\eta^2 + r^2 \mathrm{d}\varphi^2,
\end{equation}
with the Jacobian $J$ given by
\begin{equation}
  J = (\bmath{B}\bcdot \nabla \eta)^{-1}.
\end{equation}
Note that, for $B_\varphi=0$, we have $J=B^{-1}$. However, in keeping with
\citet{Hellsten79} we uphold the more general notation in order to
facilitate the inclusion of a toroidal field component in future
work. 

The contravariant components of the force balance equation
\eeref{equilibrium:force_balance} in the $\Psi$ and $\eta$ directions
read respectively
\begin{equation}
  \label{eq:equilibrium:fb_psi}
  \frac{\partial P}{\partial \Psi}+\frac{1}{J}
  \frac{\partial}{\partial \Psi}(J B^2)= A_0 N ,
\end{equation}
and
\begin{equation}
  \label{eq:equilibrium:fb_theta}
  \frac{\partial P}{\partial \eta} = A_1 N.
\end{equation}
The gravitational acceleration is directed radially inward
to the centre of the star (mass $M$). Ignoring self gravity, we have
\begin{equation}
  \label{eq:equilibrium:g}
  \bmath{g}=A_0 \bmath{e}_\Psi+A_1 \bmath{e}_\eta,
\end{equation}
with
\begin{equation}
  \label{eq:equilibrium:a0}
  A_0 = \bmath{g}\bcdot \bmath{e}_\Psi = - \frac{M}{(r^2+z^2)^{3/2}} \left( r \frac{\partial
      r}{\partial \Psi} + z \frac{\partial z}{\partial \Psi} \right),
\end{equation}
and
\begin{equation}
  \label{eq:equilibrium:a1}
  A_1 = \bmath{g} \bcdot \bmath{e}_\eta = - \frac{M}{(r^2+z^2)^{3/2}} \left( r \frac{\partial
      r}{\partial \eta} + z \frac{\partial z}{\partial \eta} \right).
\end{equation}
\citet{Payne04} took $\bmath{g}$ to be constant to simplify the
analysis, but we prefer to use Eqs. \eeref{equilibrium:a0} and
\eeref{equilibrium:a1} to allow direct comparison with the numerical
results of \citet{Payne07} and \citet{Vigelius08}.

For later comparison, we note that \eeref{equilibrium:fb_psi} and
\eeref{equilibrium:fb_theta} become formally identical to Eqs. (13) and (14) of
\citet{Hellsten79} for a stationary fluid in the absence of
gravity, rotating around the $z$ axis with constant angular velocity
$\Omega$,  if we substitute 
\begin{equation}
  \label{eq:equilibrium:centri_a0}
  A_0 = \Omega^2 r \frac{\partial r}{\partial \Psi},
\end{equation}
and
\begin{equation}
  \label{eq:equilibrium:centri_a1}
  A_1 = \Omega^2 r \frac{\partial r}{\partial \eta}.
\end{equation}
Of course, we cannot write the centrifugal force in the form
\eeref{equilibrium:g}, as it is not radial, so the substitution
\eeref{equilibrium:centri_a0} and \eeref{equilibrium:centri_a1} is
algebraic, not physical.

\subsection{Accreted magnetic mountain}
\label{sec:equilibrium:mountain}
\begin{figure*}
  \includegraphics[width=140mm,keepaspectratio]{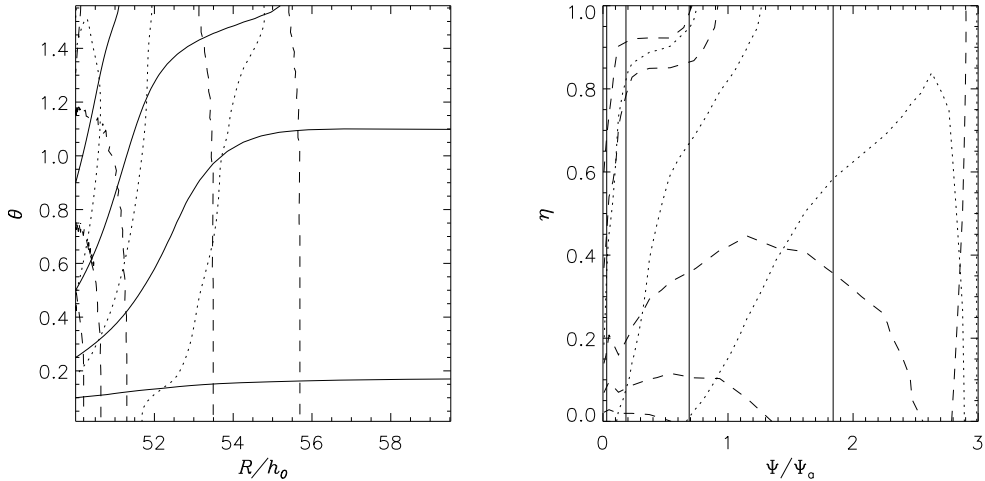}
  \caption{Hydromagnetic structure of a magnetically confined mountain
    with $M_a=M_c$ in spherical polar and field line coordinates. The left
    panel displays the density contours (dashed) $\log_{10}(N/N_0)=-13,
    -12, -11, -10.7, -10.5, -10.3$, with $N_0=5.2 \times 10^{19}$ g
    cm$^{-3}$, the magnetic field strength contours (dotted)
    $\log_{10}(B/B_0)=-7, -6, -5.5$, with $B_0=2.55 \times 10^{18}$ G,
    and the magnetic field lines (solid) with footpoints at 
    $R=R_\ast$ and $\theta=0.10, 0.12,0.15, 0.20, 0.39, 0.79$, in the
    $R$-$\theta$ plane, where $(R, \theta, \varphi)$ are standard spherical
    polar coordinates. The right panel presents the same information
    in the $\Psi$-$\eta$ plane, where $\Psi$ is normalized to the flux
    surface that closes at the inner edge of the accretion disk \change{$\Psi_a$}, and
    $\eta$ is the arc length along a field line, normalized to $0 \le
    \eta \le 1$.}
  \label{fig:equilibrium:mountain_equilibrium}
\end{figure*}

Throughout this paper, we study magnetic mountain equilibria on the
surface of a curvature-downscaled star with radius $R_\ast'=2.7 \times 10^3$ cm and mass
$M_\ast'=1.0\times10^{-5} M_\odot$. The downscaling transformation
preserves the equilibrium shape of the mountain [exactly in the small
$M_a$ limit and approximately in the large $M_a$ limit; see
\citet{Payne04, Vigelius08}], as long as the hydrostatic scale height $h_0=53.82$
cm keeps its original value for a realistic
star. Base units are $M_0=8.1\times10^{24}$ g, $N_0=5.2\times10^{19}$ g 
cm$^{-3}$, $B_0=7.2\times10^{17}$ G, and $\tau_0=5.4\times 10^{-7}$ s,
for mass, density, magnetic field, and time respectively. \change{In order to
upscale the frequencies back to a realistic neutron star, we employ
the relation $\omega^2 \propto (h_0/R_\ast)^2$
\citep{Payne06a}. While, strictly speaking, this relation only applies to
waves travelling latitudinally, we note that the field lines
are predominantly parallel to the neutron star surface
(\fref{equilibrium:mountain_equilibrium}, left panel) and the error
will be small.}

An equilibrium configuration with $M_a=M_c=1.2 \times 10^{-4} M_\odot$ is displayed in
\fref{equilibrium:mountain_equilibrium}. $M_c$ denotes the critical
accreted mass beyond which the magnetic dipole moment is substantially
reduced \citep{Payne04}. The left panel shows the
density contours (dashed) and the magnetic field lines (solid),
i.e. the flux surfaces $\Psi=\mathrm{const}$, projected into a
meridonial plane. The equilibrium configuration extends over $0 \le
\tilde{x}=(R-R_\ast)/h_0 \le 10$ and $0 \le \theta \le \pi$, where $R$
measures the radius in spherical polar coordinates,  and
$\theta$ is the colatitude. North-south symmetry is assumed.
The mountain is confined to the magnetic pole and
the distorted magnetic belt is clearly visible at $\theta \ge 0.5$ and
$r \ge 52$. The right panel shows the same plot in the $\Psi$-$\eta$
plane. Of course, the field lines are projected onto straight lines in this
plot. The density contours appear distorted since $\eta$ is
normalized to the domain $0 \le \eta \le 1$.

Contours of $B$ are plotted as dotted curves in both panels of
\fref{equilibrium:mountain_equilibrium} for $\log_{10} (B/B_0)=-7, -6,
-5.5$. The magnetic belt with its enhanced magnetic field is clearly
visible between $0.6 \le \theta \le 1.4$ in the left panel. The dotted
curves trace out isosurfaces of magnetic pressure and are therefore
useful for visualizing the Lorentz force confining the mountain.

\section{Global linear MHD oscillations}
\label{sec:linearized}
We now consider the behaviour of small-amplitude perturbations of the
magnetic mountain equilibria described in \sref{equilibrium}. The
linearized equations of ideal MHD are projected onto the field line
coordinate system in \sref{linearized:motion}. The singularities in
these equations, which determine the form of the continuous MHD
spectrum, are located in \sref{linearized:eigenvalue}.

\subsection{Equations of motion}
\label{sec:linearized:motion}
Following the notation and approach of \citet{Hellsten79}, we expand the
velocity and magnetic field perturbations in terms of their
contravariant vector components:
\begin{equation}
  \bmath{v}=v^\Psi \bmath{e}_\Psi+v^\eta \bmath{e}_\eta +
  v^\varphi \bmath{e}_\varphi,
\end{equation}
\begin{equation}
  \bmath{b}=b^\Psi \bmath{e}_\Psi+b^\eta \bmath{e}_\eta +
  b^\varphi \bmath{e}_\varphi.
\end{equation}
As in any curvilinear coordinate system, the components do not have
the same units in general, e.g. $v^\Psi$ and $v^\varphi$ have units
G cm$^2$ s$^{-1}$ and s$^{-1}$ respectively, since $\bmath{e}_\Psi$ and
$\bmath{e}_\varphi$ have units (cm G)$^{-1}$ and cm respectively.
The total pressure perturbation is defined as
\begin{equation}
  \pi = p + \bmath{B}\bcdot \bmath{b}=\change{n}+J B^2 b^\eta,
\end{equation}
where $p$ denotes the hydrostatic pressure perturbation.

We then Fourier decompose the perturbed variables with respect to
time and $\varphi$, e.g. $\bmath{v} \propto \exp[\mathrm{i}(-\omega
t+m \varphi)]$, recalling that the equilibrium is assumed to be
axisymmetric. Thus we can write down the components of
the linearized momentum balance equation,
\begin{eqnarray}
  \label{eq:linearised:l_start}
  0 & = & \frac{\omega^2 N  v^\Psi}{r^2 B^2 }-\frac{\ri \omega A_0 \change{n}x}{r^2 B^2} \nonumber \\
  & & + \ri \omega \frac{\partial
    \pi}{\partial \Psi}+\frac{\ri \omega}{J} b^\eta \frac{\partial}{\partial
    \Psi}(J^2 B^2)-\frac{\ri \omega}{J}
  \frac{\partial}{\partial \eta}\left( \frac{b^\Psi}{r^2 B^2} \right),
\end{eqnarray}
\begin{eqnarray}
  0 & = & \omega^2 N J^2 B^2 v^\eta - \ri \omega A_1 \change{n} \nonumber \\
 & &  + \ri \omega
  \frac{\partial \pi}{\partial \eta} - \ri \omega
  \frac{\partial}{\partial \eta} \left(J B^2 b^\eta \right)-\ri \omega
   b^\Psi \frac{\partial}{\partial \Psi}(J B^2),
\end{eqnarray}
\begin{eqnarray}
  0 & = & \omega^2 N r^2 v^\varphi - \omega m \pi - \frac{\ri
    \omega}{J} \frac{\partial}{\partial \eta} \left(r^2 b^\varphi \right),
\end{eqnarray}
the components of the linearized induction equation, $\partial
\bmath{b} / \partial t = \nabla \times (\bmath{v} \times \bmath{B})$, 
\begin{equation}
  \ri \omega b^\Psi + \frac{1}{J} \frac{\partial v^\Psi}{\partial \eta}=0,
\end{equation}
\begin{equation}
  \ri \omega b^\eta-\frac{1}{J} \left( \frac{\partial
      v^\Psi}{\partial \Psi}+\ri m v^\varphi\right) = 0,
\end{equation}
\begin{equation}
  \ri \omega b^\varphi+\frac{1}{J}\frac{\partial v^\varphi}{\partial
    \eta} = 0,
\end{equation}
and the linearized mass continuity equation, $\partial n/\partial t +
\nabla \bcdot (N \bmath{v})=0$,
\begin{eqnarray}
  \label{eq:linearised:l_end}
  \ri \omega n & = & \ri \omega J N b^\eta+N \frac{\partial
    v^\eta}{\partial \eta} \nonumber \\
  & & + \left( \frac{\partial N}{\partial \Psi}+\frac{N}{J}
    \frac{\partial J}{\partial \Psi}\right) v^\Psi +
  \left(\frac{\partial N}{\partial \eta}+\frac{N}{J} \frac{\partial
      J}{\partial \eta} \right) v^\eta.
\end{eqnarray}
Equations \eeref{linearised:l_start}--\eeref{linearised:l_end} are
derived by writing down the vector operators in the curvilinear
coordinate system, e.g.
\begin{equation}
  \nabla f = (r B)^2 \frac{\partial f}{\partial \Psi} \bmath{e}_\Psi +
  \frac{\partial f}{\partial \eta} \bmath{e}_\eta +
  \frac{1}{r^2} \frac{\partial f}{\partial \varphi} \bmath{e}_\varphi. 
\end{equation}

\subsection{Singular eigenvalue problem}
\label{sec:linearized:eigenvalue}
Equations \eeref{linearised:l_start}--\eeref{linearised:l_end} can
be cast into the form \citep{Hellsten79}
\begin{equation}
  \label{eq:linearised:matrix_1}
  \matrix{C} \frac{\partial \bmath{X}}{\partial \Psi}+ \matrix{D}
  \bmath{X}+\matrix{E} \bmath{Y}=0,
\end{equation}
\begin{equation}
  \label{eq:linearised:matrix_2}
  \matrix{F}\bmath{Y}+\matrix{G}\bmath{X}=0.
\end{equation}
In \eeref{linearised:matrix_1} and \eeref{linearised:matrix_2},
$\bmath{X}$ is a 2-vector with elements $v^\Psi$ and $\pi$, while
$\bmath{Y}$ is the five-vector
$(v^\eta, v^\varphi, b^\Psi, b^\eta, b^\varphi)^\mathrm{T}$.
$\matrix{C}$ and $\matrix{D}$ are matrices, while $\matrix{E}$, $\matrix{F}$, and
$\matrix{G}$ are differential matrix operators involving ordinary
derivatives in $\eta$. Using \eeref{linearised:matrix_2},
$\bmath{Y}$ can be eliminated from \eeref{linearised:matrix_1} to give
\begin{equation}
  \label{eq:linearised:master}
  \frac{\partial \bmath{X}}{\partial \Psi}+\matrix{C}^{-1}
  (\matrix{D}-\matrix{E} \matrix{F}^{-1} \matrix{G} ) \bmath{X}=0.
\end{equation}

The continuous part of the frequency spectrum consists of the values of
$\omega$ for which \eeref{linearised:master} becomes singular at a
flux surface. This happens when either
\begin{equation}
  \label{eq:linearised:flow}
  \matrix{C}\bmath{X} = 0
\end{equation}
or
\begin{equation}
  \label{eq:linearised:other}
  \matrix{F}\bmath{Y} = 0
\end{equation}
have nontrivial solutions at some $\Psi=\Psi_0$. The flow continuum,
\eref{linearised:flow}, has only the trivial eigenvalue
$\omega=0$. Equation \eeref{linearised:other} defines the Alfv\'{e}n and cusp
continua. It can be rewritten in the form \citep{Hellsten79}
\begin{equation}
  \label{eq:linearised:eigenvalue}
  \omega^2 \matrix{P} \bmath{\eta} + \matrix{L} \bmath{\eta} = 0,
\end{equation}
where $\bmath{\eta}=(J v^\eta$, $v^\varphi)^\mathrm{T}$
and the matrices $\matrix{P}$ and $\matrix{L}$ are given by
\begin{equation}
  \label{eq:linearised:p}
  \matrix{P} = N \left(
    \begin{array}{cc}
      B^2 & 0 \\
      0 & r^2\\
    \end{array}
    \right),
\end{equation}
\begin{equation}
  \matrix{L} = \left[
  \label{eq:linearised:l}
    \begin{array}{ccc}
      \bmath{B}\bcdot \nabla  \left( \frac{P B^2}{B^2+P} \bmath{B}\bcdot
      \nabla \right)+ A & 0
      \\
      0 & \bmath{B}\bcdot \nabla \left(r^2 \bmath{B} \bcdot \nabla \right)
    \end{array}
  \right],
\end{equation}
with
\begin{equation}
  A=\bmath{B}\bcdot \nabla \left(\frac{B^3 N A_1}{B^2+P}\right)  - B A_1
  \bmath{B}\bcdot \nabla N+\frac{B^2 N^2 A_1^2 }{B^2 + P}.
\end{equation}

The eigenvalue  \eref{linearised:eigenvalue}  decouples
into two  second-order, ordinary differential equations (involving
derivatives $\bmath{B} \bcdot \nabla = B \mathrm{d}/\mathrm{d}\eta$), one
for the Alfv\'{e}n continuum and one for the cusp
continuum. The problem is self-adjoint \citep{Hellsten79}, so the
eigenvalues $\omega^2$ are real.
In the absence of gravity ($A=0$), the operator $\matrix{L}$ is
negative and both continua are real ($\omega^2 \ge 0$). Gravity
provides a constant offset $A$, such that the Alfv\'{e}n continuum is
stable while the cusp continuum may be unstable. We note that the cusp
continuum is the two-dimensional equivalent of the slow magnetosonic
continuum in a one-dimensional, gravitating plasma slab (see
\sref{spectrum:eigenfunctions} and references therein for a full discussion).

\section{Continuous spectrum of a magnetic mountain}
\label{sec:spectrum}

\subsection{Algorithm}
We solve \eref{linearised:eigenvalue} numerically for the continuous
spectrum by following the procedure below.
\begin{enumerate}
\item We compute the axisymmetric
Grad-Shafranov equilibrium, employing an iterative relaxation
algorithm developed by \citet{Payne04} (see
\sref{equilibrium:mountain}).
\item The magnetic field equation,
  $\mathrm{d}\bmath{r}/\mathrm{d}\eta=\bmath{B}(\bmath{r})$, is
integrated via a fourth order Runge-Kutta algorithm \citep{Press86} to
obtain field lines $r=r(\eta, \Psi_0)$, $z=z(\eta, \Psi_0)$ starting from
different footpoints $\Psi_0$ at the stellar surface. 
\item The field values and their derivatives are evaluated along
  the field lines. The entries in $\matrix{P}$ and $\matrix{L}$ are
  computed by spline interpolation in $\eta$. 
\item We find $\omega^2$ and the
associated eigenfunction by using a shooting algorithm \citep{Press86}
to integrate \eeref{linearised:eigenvalue} separately for the
Alfv\'{e}n and cusp continua. The integrator is  fourth-order Runge-Kutta
with adaptive step size control.
\end{enumerate}

 The boundary conditions for the eigenfunctions along any field line
($\Psi_0,\varphi_0$), are:
\begin{displaymath}
\mathrm{(i)}\;J v^\eta (\eta=0)=v^\varphi(\eta=0)=0,\; v^\eta(\eta=1)=v^\varphi(\eta=1)=0
\end{displaymath}
when the field line is closed, or
\begin{displaymath}
\mathrm{(ii)}\;\partial_\eta (J v^\eta) (\eta=1)=\partial_\eta
v^\varphi(\eta=1)=0   
\end{displaymath}
when the field line leaves the integration
area, consistent with \citet{Payne07} and \citet{Vigelius08}. The boundary
conditions at the stellar surface enforce line tying, while the
zero-gradient outer boundary condition crudely approximates the
magnetosphere-accretion disk coupling [cf. the discussion in
\citet{Vigelius08}].

We present the eigenfunctions $J v^\eta$ and $v^\varphi$ for four
selected fieldlines in section \ref{sec:spectrum:eigenfunctions}, two
leaving the integration volume and two returning to the surface. The
continuous spectrum and its dependence on the accreted mass $M_a$ and
surface magnetic field strength $B_\ast$ are discussed in section
\ref{sec:spectrum:eigenvalues}. 

\subsection{Eigenfunctions}
\label{sec:spectrum:eigenfunctions}
We preface this subsection by briefly reviewing the physical origin of
the continuous spectrum in a plane-parallel, gravitating (but not
self-gravitating)  plasma slab. This analogous system can be treated
analytically and is helpful when interpreting the results for a
magnetic mountain. Here, we follow the exposition in
\citet{Goedbloed04}.

Consider an infinite slab in Cartesian
coordinates $(x,y,z)$, whose magnetic flux surfaces are perpendicular to the
gravitational acceleration (directed along the $x$-axis). We render
the problem one-dimensional by assuming that all equilibrium
quantities $N(x)$ and $\bmath{B}(x)=B(x) \bmath{e}_y$ depend only on the height $x$ or,
equivalently, on $\Psi(x)$.
In this case, \eref{linearised:master} exhibits a genuine
singularity, i.e. the associated eigenfunctions become singular, when
the eigenvalue $\omega^2$ equals the local Alfv\'{e}n 
or slow magnetosonic frequency, i.e. either
$\omega^2=\omega^2_\mathrm{A}(x)=(\bmath{k}_0 \bcdot \bmath{B})^2/N$ or
$\omega^2=\omega^2_\mathrm{S}(x)=N \omega^2_\mathrm{A}(x)/(N+B^2)$.
The horizontal wave vector $\bmath{k}_0$ is the projection of $\bmath{k}$
onto the magnetic flux surface. All eigenfunctions can be written as
$\xi(\bmath{x})=\xi_0(\Psi) \exp [\mathrm{i} (k_y y+k_z z)]$, with $k_0
= (k_y^2+k_z^2)^{1/2}$.  It can be shown that $v^\Psi$ is then
square integrable, involving a logarithmic singularity, while $v^z$ is
not square integrable 
for the Alfv\'{e}n continuum, and $v^y$ is not square integrable for
the slow modes. In addition,
\eeref{linearised:master} exhibits an apparent singularity, when
$\omega^2$ equals the local magnetosonic turning point
frequencies, i.e. the slow and fast magnetosonic frequencies for
$k_x=0$. In this case, the eigenfunctions remain finite\footnote{An
  even simpler system is obtained by considering an exponentially 
stratified atmosphere with uniform sound and Alfv\'{e}n
speeds, i.e. constant $\omega^2_\mathrm{A}$ and $\omega^2_\mathrm{S}$.
Under these circumstances, the continuous spectra degenerate into a single
point each, which are cluster points of the discrete
spectra. This case includes the Parker instability
\citep{Parker1967, Mouschovias74}. 
The Parker instability is responsible for the
transient, three-dimensional, ideal-MHD relaxation of an initially
axisymmetric, magnetically confined mountain observed in previous
numerical simulations \citep{Vigelius08}.}.

\change{The mountain equilibrium exhibits logarithmic singularities in
$\bmath{X}$ and $\bmath{Y}$ for the Alfv\'{e}n and cusp
continuum \citet{Hellsten79}. Naturally, in a realistic scenario perturbations are damped
by nonideal effects such as viscosity and resistivity and singular
eigenfunctions cannot arise. For example,
resistivity will limit the maximum fractional amplitude of a magnetic
perturbation to $\delta B/B \le \ln [(\tau_\mathrm{C} \eta)^{1/2}]$,
where $\tau_\mathrm{C}$ is a characteristic time scale and $\eta$ is the
resistivity. If we choose $\tau_\mathrm{C}=\tau_\mathrm{A}=2.2 \times 10^{-2}$ cm
s$^{-1}$ and $\eta=1.3 \times 10^{27}$ \citep{Vigelius08c} we find
$\delta B/B \la 30$. Of course, the linear approximation breaks down
at such a high fractional amplitude.}

\begin{figure*}
  \includegraphics[width=140mm,keepaspectratio]{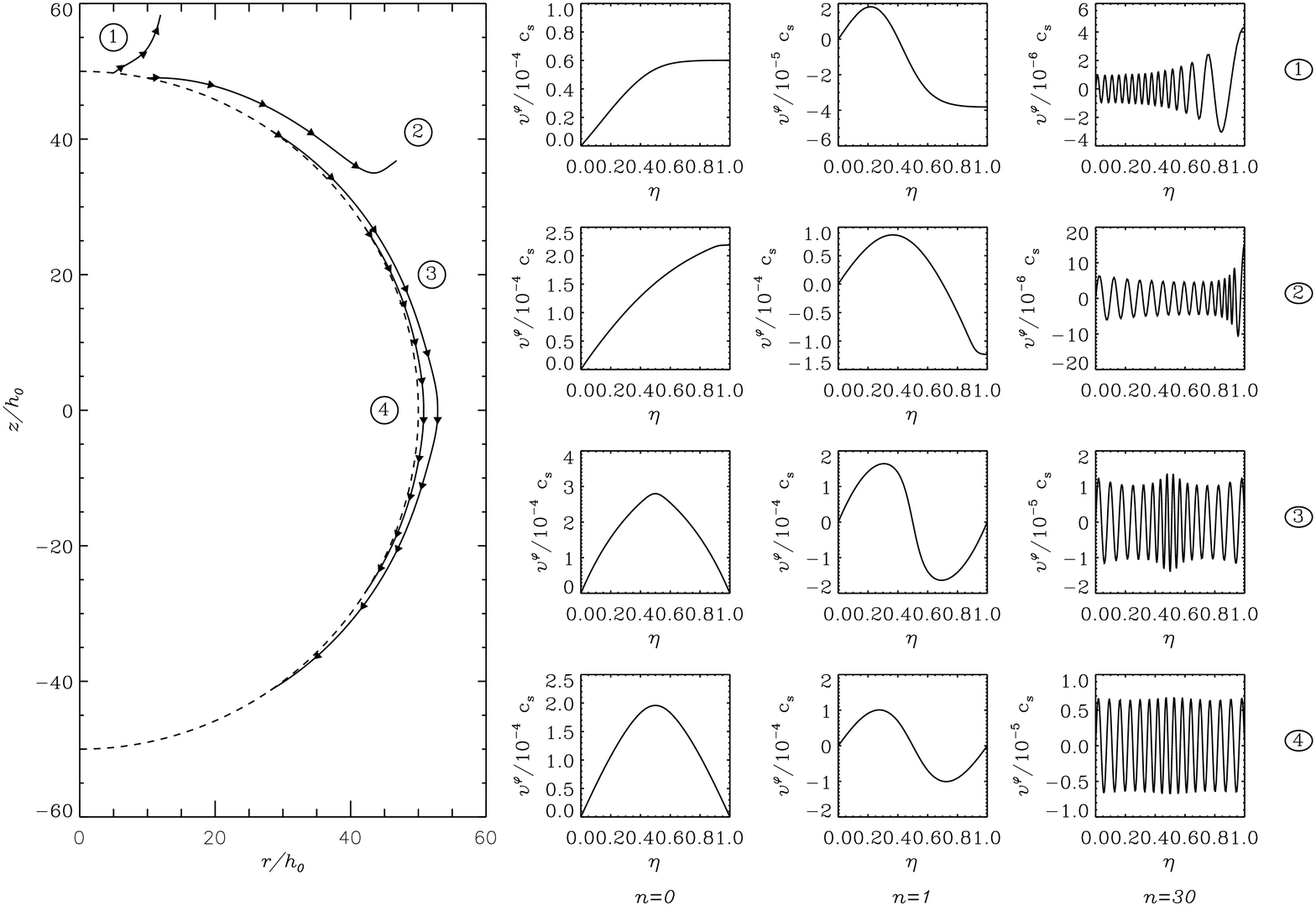}
  \caption{Toroidal velocity eigenfunctions $v^\varphi$ of the Alfv\'{e}n continuum for
    a magnetic mountain with accreted mass $M_a=M_c=1.2 \times 10^{-4}
    M_\odot$. The foot points of the field lines \flone, \fltwo,
    \flthree, and \flfour\ are at $\theta=0.1,
    0.2, 0.6, 1$ rad (left panel, tracing the field lines of the
    mountain in the meridional plane). The right panels graph
    $v^\varphi$ against arc length, $\eta$,
    along the four field lines (top to bottom) for node numbers
    $n=0$ (left), 1 (middle), 30 (right). $h_0=53.82$ cm is the
    characteristic scale height (units of axes in left panel) 
    and $c_s=10^8$ cm s$^{-1}$ is the sound speed. $\eta$ is
    normalized to the length of the field line, such that $0 \le \eta \le 1$.}
  \label{fig:spectrum:alfven_eigen}
\end{figure*}
We are now ready to apply these ideas to the continuous spectrum of a
two-dimensional, magnetically confined mountain. In a two-dimensional system, all eigenfunctions are functions of
$\Psi$ and $\eta$. \fref{spectrum:alfven_eigen} (right panels) shows the eigenfunctions
$v^\varphi(\eta)$ for the Alfv\'{e}n continuum [lower row of
Eq. \eeref{linearised:eigenvalue}] of an accreted mountain
with $M_a=M_c$. The foot points of the field
lines are at $\theta=0.1, 0.2, 0.6, 1$ rad, corresponding to
$\Psi_0/\Psi_a=0.03, 0.12, 0.96, 2.12$. \change{Here, $\Psi_a$ is the flux
    surface that closes at the inner edge of the accretion disk.}
The field lines are traced out
and labelled in the left
panel. The right panels show $v^\varphi(\eta)$ for the same field
lines (top to bottom)
for different numbers $n=0, 1, 30$, where $n$ equals the number
of nodes in the eigenfunction.

Field lines \flone\ and \fltwo\ leave
the integration volume, while field lines \flthree\ and \flfour\ close back onto
the surface. Field line \flone\ exhibits only small
curvature. Consequently, $v^\varphi$ is sinusoidal. Its envelope
increases exponentially $\propto \exp(1.5 \eta)$ (best seen in the rightmost
panel). Its wavelength also increases with $\eta$, which can be understood by
noting that the Alfv\'{e}n speed $v_\mathrm{A} \propto \exp(4.5 \eta)$ increases towards
the outer boundary. In contrast, field line \fltwo\ runs parallel to the
stellar surface until it takes a sharp turn at $\eta \approx
0.8$. Thus, $v^\varphi$ is a sine wave whose amplitude and wavelength
remain constant until the curvature term $\bmath{B}\cdot\nabla$
changes at the bend, sharply increasing the amplitude and the
wavelength.

The closed field lines \flthree\ and \flfour\ behave like field line
\fltwo. Line \flthree\ shows a spiky 
feature at $\eta=0.5$ (equator) due to the change of curvature
locally. Line \flfour\ is almost a perfect sine wave for all
$\eta$. Recall that, in a plane-parallel slab, $N$ and $B$ (and
consequently $v_\mathrm{A}$) remain constant for all $\eta$. The
eigenfunctions are hence pure sine waves with the dispersion relation
$v_\mathrm{A}=\omega/k$. We discuss this dispersion relation
quantitatively in the next section.

\begin{figure*}
  \includegraphics[width=140mm,keepaspectratio]{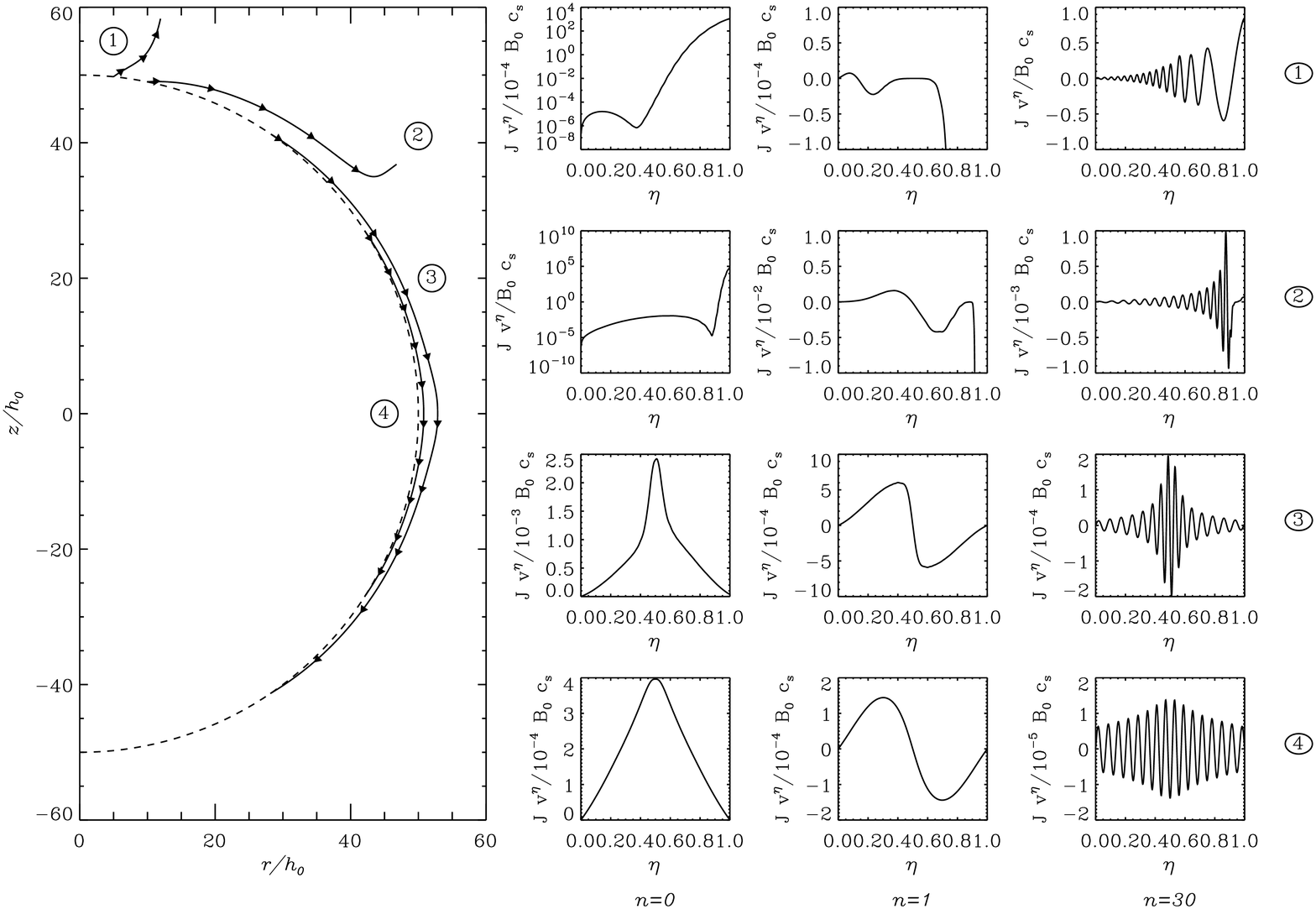}
  \caption{Field-aligned velocity eigenfunctions $J v^\eta$ of the cusp continuum for
    a magnetic mountain with accreted mass $M_a=M_c=1.2 \times 10^{-4}
    M_\odot$. The foot points of the field lines \flone, \fltwo,
    \flthree, and \flfour\ are at $\theta=0.1,
    0.2, 0.6, 1$ rad (left panel, tracing the field lines of the
    mountain in the meridional plane). The right panels graph
    $J v^\eta$ against arc length, $\eta$,
    along the four field lines (top to bottom) for node numbers
    $n=0$ (left), 1 (middle), 30 (right). $h_0=53.82$ cm is the
    characteristic scale height (units of axes in left panel)
    and $c_s=10^8$ cm s$^{-1}$ is the sound speed. $\eta$ is
    normalized to the length of the field line, such that $0 \le \eta \le 1$.}
  \label{fig:spectrum:cusp_eigen}
\end{figure*}
The eigenfunctions $J v^\eta$ of the cusp continuum, displayed in
\fref{spectrum:cusp_eigen} (right panels), behave similarly. Field
lines \flone\ and \fltwo\ run parallel to the stellar surface until they bend radially
outwards. Along these field lines, $J v^\eta$ is essentially sinusoidal, with a sharp rise in
amplitude at $\eta \approx 0.4$ and $\eta \approx 0.8$
respectively. This behaviour can be explained by the presence of the
coefficient $N B^2/(B^2+N)$, which contains the
partial hydrodynamic pressure (the factor $B^2$ comes in since we
compute $J v^\eta$ instead of $v^\eta$), which is absent in the
equations for the Alfv\'{e}n continuum. A comparison with
\fref{equilibrium:mountain_equilibrium} shows that $B$ and $N$
flatten out considerably at $\eta=0.4$ ($\eta=0.8$) for field line
\flone\ (\fltwo). Consequently, the derivative of the coefficient is
steep up to that point and becomes negligible thereafter. Field lines
\flthree\ and \flfour\ behave essentially like their Alfv\'{e}n
counterparts, showing the same sharp spike at $\eta=0.5$ as 
lines \flone\ and \fltwo. Remember that the field lines \flthree\ and
\flfour\  cover only one hemisphere, i.e. $\eta=1$ corresponds to the
equator (see \fref{equilibrium:mountain_equilibrium}).

We summarize our results so far. The eigenfunctions for the
Alfv\'{e}n continuum are sine waves with an exponential envelope. The
wave length increases with $\eta$ due to the increase in the
Alfv\'{e}n speed. The eigenfunctions of the cusp continuum show a
characteristic spike at the point where the field lines point radially
outwards (and, less distinctly, at the equator for the closed field
lines). This spike is a consequence of the flattening out of the
$\eta$ profile of the partial hydrodynamic pressure.

\subsection{Eigenvalues}
\label{sec:spectrum:eigenvalues}
We now evaluate the frequency spectrum for the magnetic mountain in
Figs. \ref{fig:spectrum:alfven_eigen} and
\ref{fig:spectrum:cusp_eigen} by computing $\omega^2$ for polar (footpoint at $\theta=0.01$
rad) and equatorial (footpoint at $\theta=1.6$ rad) field lines. Since
$B(R=R_\ast)$ decreases monotonically with $\Psi$, these two field lines bracket the continuous spectrum of the whole configuration. 
The results in \sref{spectrum:eigenfunctions} for the four 
field lines in Figs. \ref{fig:spectrum:alfven_eigen} and
\ref{fig:spectrum:cusp_eigen} do not show any evidence for a spectrum that folds over
onto itself, a feature in some other MHD systems [cf. the
discussion in \citet{Goedbloed04}]. The spectrum folds onto
itself when the characteristic speeds (Alfv\'{e}n and slow-magnetosonic in
the plane parallel slab) have local extrema in the $\Psi$ range considered.

\begin{figure}
  \includegraphics[width=84mm,keepaspectratio]{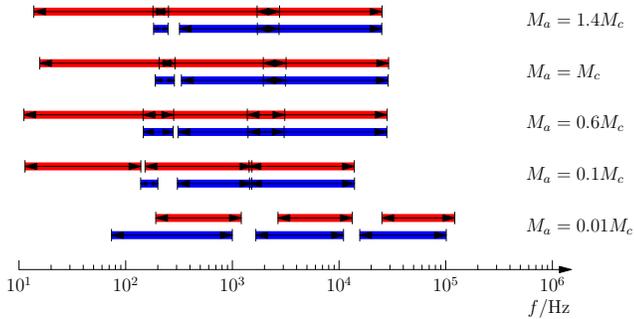}
  \caption{Alfv\'{e}n (red bands) and cusp (blue bands) continuous
    frequency spectrum (in Hz) for
    magnetic mountains with (from top to bottom) $M_a/M_c=1.4, 1, 0.6, 0.1,
    0.01$, computed for the node numbers (left bands to right bands) $n=0, 10, 100$.}
  \label{fig:spectrum:mass_variation}
\end{figure}
\fref{spectrum:mass_variation} displays $\omega^2$ for $M_a/M_c=1.4,
1, 0.6, 0.1, 0.01$. We upscale all frequencies to a realistic
neutron star according to $\omega^2 \propto (h_0/R_\ast)^2$ \citep{Vigelius08}. The
Alfv\'{e}n (cusp) continuum is drawn in red (blue), as a
shaded interval with double-headed arrows, for node numbers $n=0, 10,
100$. We stress that the gaps appearing in 
\fref{spectrum:mass_variation} stem from our choices of
$n$; in fact, the 
whole spectrum $\omega^2 \ge \omega^2_\mathrm{min}$ is covered without
gaps by nodes $n\ge 0$. Furthermore, both the Alfv\'{e}n and the cusp continua
for different $n$ overlap. Overlap is common in some MHD systems,
where multiple degeneracies of the eigenfunctions occur.


Let us compare the results in \fref{spectrum:mass_variation} to the frequency spectrum of a
plane parallel slab.  We remind the reader that, in a one-dimensional system like the
plane-parallel slab, the spectra for 
different $n$ are defined by the dispersion relations given in
paragraph two of \sref{spectrum:eigenfunctions}, since the equilibrium values
and thus the coefficients of \eref{linearised:eigenvalue} do not
depend on $\eta$. The lower bound for the Alfv\'{e}n continuum is
measured from \fref{spectrum:mass_variation} to be
$\omega_\mathrm{A,low}/(2 \pi\,\mathrm{Hz})=15.6$ for $M_a/M_c=1$.
At the same time, the local Alfv\'{e}n frequency for the polar field
line, lies in the range
$3.39 \le \omega_\mathrm{A}/(2 \pi\,\mathrm{Hz})\, \le 55.38$. Our aim is to
pick the $n=0$ mode, so we assume $k_0 \approx 1/h_0$. In our two-dimensional system, $\omega^2_\mathrm{S}$
and $\omega^2_\mathrm{A}$ are both functions of $\eta$. The first grid
cell is centered at $\theta=0.01$ rad and hence determines the lowest
polar field line that we can choose. This field line runs almost
parallel to the $z$ axis, i.e. $r\approx \mathrm{const}$. Hence, the
bottom row of \eeref{linearised:eigenvalue} reduces to the dispersion
relation for Alfv\'{e}n waves. Consequently,
$\omega_\mathrm{A,low}$ lies in the range of the local Alfv\'{e}n
frequencies. On the other hand, we can write down the dispersion
relation for the cusp continuum assuming that $N$ and $B$ do not
depend on $\eta$, viz. $\omega^2 = \omega_\mathrm{S}^2-A/(N B^2)$.
The singular frequencies are offset by a constant due to the
gravitational force. We compute $\omega(\eta)$ for the polar field line to be
$154 \le \omega/(2 \pi\,\mathrm{Hz})\le 366$.  Again, we find that
$\omega_\mathrm{C,low}/(2 \pi\,\mathrm{Hz})=189$ (cusp continuum) lies in this range.

For completeness, we display
$\omega$ for the four field lines in \sref{spectrum:eigenfunctions} as
a function of the node number $n$ in
\fref{spectrum:lines_spectrum_zeus.eps}. The Alfv\'{e}n (cusp)
eigenfrequencies are plotted as plus (star) symbols. For field lines
\flone\ and \fltwo, we find $\omega_\mathrm{C} >
\omega_\mathrm{A}$, consistent with the discussion in the previous
paragraph. However, field lines \flthree\ and \flfour\ obey $\omega_\mathrm{A} >
\omega_\mathrm{C}$. The reason is that the latter fieldlines stay
close to the surface. As a consequence, the constant offset, which is
set by the gravitational force, is on average $\approx 2.5$ times
lower for \flthree\ and \flfour, compared to \flone\ and \fltwo. We
find $\omega_\mathrm{A} \rightarrow \omega_\mathrm{C}$ for
$n\rightarrow 0$ for \flthree\ and \flfour\ , because $A$ is neglible
in this case and the eigenfrequencies depend on the node number as
$\omega^2 \propto n^2$ with different coefficients for
$\omega_\mathrm{A}$ and $\omega_\mathrm{C}$.

\begin{figure}
  \includegraphics[width=84mm,keepaspectratio]{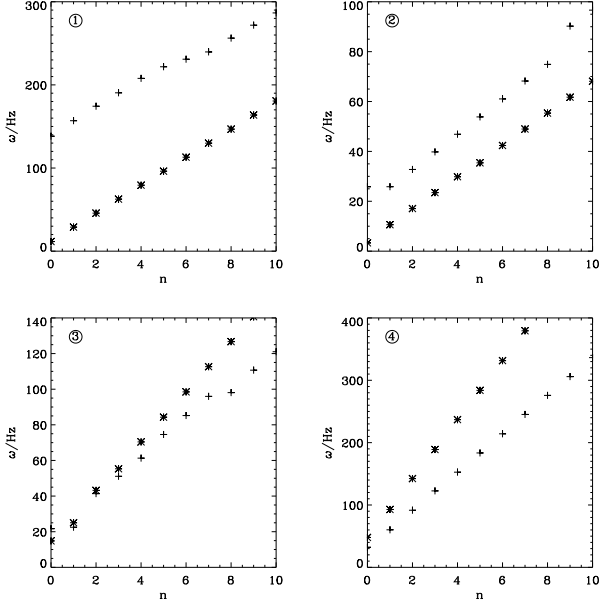}
  \caption{Alfv\'{e}n (star) and cusp (plus) eigenfrequencies for a
    magnetic mountain with $M_a=M_c$ and $B_\ast=10^{12}$ G for the
    four field lines \flone--\flfour\ in
    \sref{spectrum:eigenfunctions} (four panels, labelled at top left)
    as a function of node number $0\le n \le 10$.} 
  \label{fig:spectrum:lines_spectrum_zeus.eps}
\end{figure}

We are now in a position to explore the dependence of
$\omega_\mathrm{low}$ on the accreted mass $M_a$
(\fref{spectrum:mass_variation}) and on the surface magnetic field
$B_\ast$ (\fref{spectrum:mag_variation}). We do this by computing the
Alfv\'{e}n and cusp frequency ranges employing the dispersion
relations for constant field values established in the previous
paragraphs and examine how they change with $M_a$ and $B_\ast$,
respectively. The results are tabulated in table \ref{tab:frequencies}.

\begin{table*}
  \centering
  \caption{Comparison of the lower bounds of the Alfv\'{e}n and cusp
    frequency spectra with the range computed under the assumption of
    constant field values (see text in
    \sref{spectrum:eigenvalues}). The top and bottom panels record the
    eigenfrequencies versus accreted mass $M_a$ and surface magnetic
    field $B_\ast$ respectively.}
  \begin{tabular}{@{}ccccc}
    \hline
    $M_a/M_c$ & $\omega_\mathrm{A,low}/2 \pi$ & $\omega_\mathrm{C,low}/2\pi$ &
    Alfv\'{e}n range & cusp range \\
     & [Hz] & [Hz] & [Hz] & [Hz] \\
     \hline
     0.01 & 191 & 74 & $17.9-1537$ & $37.9-390$\\
     0.1 & 11.4 & 200 & $1.76-65.7$ & $158 - 519$ \\
     0.6 & 11.1 & 146 & $2.17-33.3$ & $40.3-481$ \\
     1   & 16 & 189 & $2.81-55.4$ & $99.1-483$  \\
     1.4 & 14 & 183 & $2.50-45.7$ & $26.8-474$ \\
    \hline
    $B_\ast/10^{12}\mathrm{G}$ & $\omega_\mathrm{A,low}/2\pi$ & $\omega_\mathrm{C,low}/2\pi$ &
    Alfv\'{e}n range & cusp range \\
     & $[\mathrm{Hz}]$ & $[\mathrm{Hz}]$ & [Hz] & [Hz] \\
    \hline
    0.1 & 11.8 & 164 & $2.57-40.5$ & $101-368$ \\
    1   & 13.8 & 177 & $2.57-40.5$ & $101-368$ \\
    10  & 13.8 & 177 & $2.57-40.5$ & $101-368$ \\
    \hline
\end{tabular}
  \label{tab:frequencies}
\end{table*}

The lower bounds for the
Alfv\'{e}n (cusp) continuum are found to be
$\omega_\mathrm{A,low}/(2 \pi\,\mathrm{Hz})=191, 15.6, 13.7$
[$\omega_\mathrm{C,low}/(2 \pi\,\mathrm{Hz})=73.7, 189, 183$] for 
$M_a/M_c=0.01, 1, 1.4$, respectively. It is interesting to compare
$M_a=0.01 M_c$ (early stage of magnetic burial) with $M_a=M_c$
(middle stage). For $M_a=0.01 M_c$, the mass
density is $\approx 0.03$ times the value at $M_a=M_c$ and we find
$\bar{B}(M_a=0.01 M_c) \approx 2 \bar{B}(M_a = 
M_c)$, where $\bar{B}$ is averaged over field line \flone. The
gravitational term $A$ is
only moderately affected, with $\bar{A}(M_a=0.01 M_c) \approx 
0.5 \bar{A}(M_a=M_c)$. On the other hand, we
find $\bar{B}(M_a=1.4 M_c)\approx 0.95 \bar{B}(M_a=M_c)$, $\bar{N}(M_a=1.4
M_c)\approx 1.24 \bar{N}(M_a=M_c)$, and $\bar{A}(M_a=1.4 M_c)\approx 1.1
\bar{A}(M_a=M_c)$. This is the reason why $\omega_\mathrm{low}$ peaks at
$M=M_c$. For $M=1.4 M_c$, the magnetic field strength remains
unchanged but the additional accreted matter increases $N$.

\begin{figure}
  \includegraphics[width=84mm,keepaspectratio]{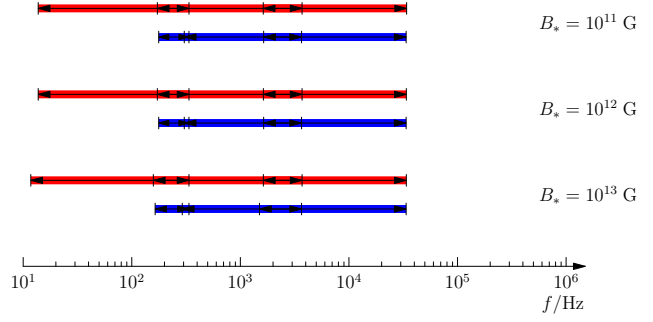}
  \caption{Alfv\'{e}n (red bands) and cusp (blue bands) continuous
    frequency spectrum (in Hz) for 
    magnetic mountains with (from top to bottom) polar magnetic field
    strength $B_\ast/\mathrm{G}=10^{13}, 10^{12}, 10^{11}$, computed
    for the node numbers (left bands to right bands) $n=0, 10, 100$.} 
  \label{fig:spectrum:mag_variation}
\end{figure}
We redo the analysis of the previous paragraph, this time varying the polar magnetic
field $B_\ast$ (\fref{spectrum:mag_variation}). The lower bounds for the
Alfv\'{e}n (cusp) continuum are
$\omega_\mathrm{A,low}/(2 \pi\,\mathrm{Hz})=11.7,13.8,13.8 $
[$\omega_\mathrm{C,low}/(2 \pi\,\mathrm{Hz})=164,177,177$] for 
$B_\ast/\mathrm{G}=10^{11}, 10^{12}, 10^{13}$ and $M_a=M_c$
respectively. The continuous spectrum does not shift much as $B_\ast$
varies over two decades. This is not surprising: $M_c$
scales as $M_c \propto B_\ast^2$, so that two mountains with a
different $B_\ast$ but the same $M_a/M_c$ effectively share the same
steady-state hydromagnetic structure and hence the same MHD spectrum. 

\subsection{Rotational splitting}
\label{sec:spectrum:splitting}
One expects to find magnetically confined mountains on rapidly
rotating neutron stars with $0.1\,\mathrm{kHz}\la\Omega/2 \pi \la
0.7\,\mathrm{kHz}$, e.g. in low-mass X-ray binaries, and accreting
millisecond pulsars. In such objects, the Coriolis force is an
important factor in determining stability. We do not treat the
Coriolis force in this article, because our main aim is to compare the
analytically derived MHD spectrum with the output from
\textsc{zeus-mp} simulations in the literature, which set $\Omega=0$
\citep{Payne07, Vigelius08}. However, the framework in
\sref{linearized} and \citet{Hellsten79} can easily accomodate $\Omega
\ne 0$, as foreshadowed at the end of \sref{linearized:motion}.

Rotation splits the Alfv\'{e}n and cusp continua,
possibly destabilizing them. We perform an order-of-magnitude calculation to estimate the
influence of rapid rotation on stability. Note first that the constant
term $A$ due to gravitation exceeds its centrifugal counterpart
($A_\mathrm{1, grav}\sim G M_\ast/R_\ast \gg A_\mathrm{1,
  rot}\sim \Omega^2 R_\ast$), justifiying the neglect of the centrifugal force
in previous numerical papers \citep{Payne07, Vigelius08}. For $\Omega
\ne 0$, the two equations of \eeref{linearised:eigenvalue} couple via
the Coriolis force \citep{Hellsten79}:
\begin{eqnarray}
  \label{eq:discussion:oom1}
  0 & = & \sigma^2 N B^2 J v^\theta  - \i 2 \sigma \Omega N r (\bmath{B} \bcdot
  \nabla r) v^\varphi + \nonumber \\
 & & \bmath{B} \bcdot \nabla \frac{p
    B^2}{B^2+\gamma p} \bmath{B}\bcdot \nabla J v^\theta,
\end{eqnarray}
and
\begin{eqnarray}
  \label{eq:discussion:oom2}
  0=\sigma^2 N r^2 v^\varphi +  \i 2 \sigma \Omega N r (\bmath{B} \bcdot
  \nabla r) J v^\theta + \bmath{B} \bcdot \nabla r^2 \bmath{B} \bcdot
  \nabla v^\varphi.
\end{eqnarray}
\change{Clearly, the Coriolis force, which produces the second term in each
equation, dominates the other terms.
For \eeref{discussion:oom1}, we find $| [2 \sigma \Omega N r (\bmath{B}\bcdot \nabla
      r)]/(\sigma^2 N B) | \sim \Omega/\sigma = 10^{2} \gg 1$,
and for the third term $| [2 \sigma \Omega N r (\bmath{B}\bcdot \nabla
    r)]/[\bmath{B} \bcdot \nabla (N B^2)(B^2+\gamma p) \bmath{B}
    \bcdot \nabla J] | \sim (\sigma \Omega \eta^2)/c_s ^2
  =10^6 \gg 1$.
Equivalently, for \eeref{discussion:oom2}, we find $| [2 \sigma \Omega N r (\bmath{B}\bcdot \nabla
      r J)]/(\sigma^2 N r^2)| \sim \Omega/\sigma = 10^{2} \gg 1$,
and $| [2 \sigma \Omega N r (\bmath{B}\bcdot \nabla
      r J)]/(\bmath{B} \bcdot \nabla r^2 \bmath{B} \bcdot \nabla)
  | \sim \sigma \Omega N/B^2 = 10^{6} \gg 1$.}  
A more general analysis including rotation is therefore needed in the future.

\section{Discussion}
\label{sec:discussion}

In this article, we compute semi-analytically the continuous part of
the ideal-MHD frequency spectrum of axisymmetric, magnetically confined
mountains. We find that the continuous spectrum covers all frequencies
above a minimum
$\omega_\mathrm{min}$. Furthermore, we find $\omega^2_\mathrm{min} > 0$ in all
the configurations ($0.01 \le M_a/M_c \le 1.4$) we study. This
further substantiates two important properties deduced previously from
numerical simulations: (i) magnetic mountains are marginally stable,
i.e., $\omega$ has zero imaginary part; and (ii) mountains relax hydromagnetically
through the undulating submode of the three-dimensional Parker
instability \citep{Vigelius08}, which possesses discrete
eigenvalues only. We also find that, for  $M_a\ga M_c$, when the magnetic structure of
the mountain substantially deviates from a dipole, $\omega_\mathrm{min}$
for the Alfv\'{e}n and cusp continua depends weakly on $M_a$ and $B_\ast$.

\begin{figure}
  \includegraphics[width=84mm,keepaspectratio]{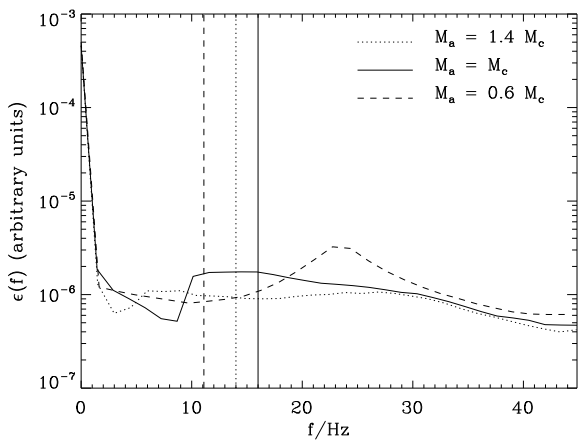}
  \caption{Fourier transform $\epsilon(f)$ of the mass ellipticity
  taken from \textsc{zeus-mp} simulations of magnetic mountains with
  $M_a=M_c$ (solid), $M_a=1.4 M_c$ (dotted), and $M_a=0.6 M_c$
  (dashed). The continuous Alfv\'{e}n spectrum is clearly
  visible for $M_a=M_c$ as a band starting at $\approx 11$ Hz. We mark
  the theoretical analytic values of $f_\mathrm{A,low}=11.1, 16, 14\;
  \mathrm{Hz}$ for $M_a/M_c=0.6, 1, 1.4$, respectively, as 
  vertical lines.  Note that the cusp continuum, with
  $\omega_\mathrm{C,low}=177.07$ Hz, lies outside the range of this plot,
  since it exceeds the Nyquist limiting frequency
  $\omega_\mathrm{N}=44.5$ Hz.}  
  \label{fig:spectrum:spectrum_comparison}
\end{figure}
How do our analytic results compare with numerical simulations of
oscillating magnetic mountains \citep{Payne07, Vigelius08}? To answer
this question, we perform an axisymmetric ideal-MHD
simulation of a magnetic mountain with $M_a=M_c$, which is perturbed
slightly at $t=0$. The simulation is
performed using the parallel ideal-MHD solver \textsc{zeus-mp}
\citep{Hayes06}. \fref{spectrum:spectrum_comparison} shows the Fourier
transform $\epsilon(f)$ of the mass ellipticity $\epsilon(t)$, which is
proportional to the mass quadrupole moment of the mountain [see
Eq. (2) in \citet{Vigelius08}]. We choose $\epsilon$ to compute the spectrum as (i) it
is directly measurable from future gravitational wave data
\citep{Vigelius08b} and (ii) it is an integrated
quantity sampling $N$ and $\bmath{B}$ everywhere, so it is sensitive
to all global oscillation modes.

Besides the constant offset at $f=0$, the simulation with $M_a=M_c$
(solid curve in \fref{spectrum:spectrum_comparison}) exhibits a continuous spectrum
with a lower boundary at $f \approx 11\; \mathrm{Hz}$. This is
consistent with the analytic theory in \sref{spectrum}, which yields
$f_\mathrm{A,low}=16\; \mathrm{Hz}$, indicated as a vertical line in
\fref{spectrum:spectrum_comparison}. The width of a frequency 
bin is $\delta f=1.45$ Hz, so we conclude that the lower boundary of
the simulated spectrum almost coincides with $f_\mathrm{A,low}$. It
appears that the lowest modes of the continuous
Alfv\'{e}n spectrum are indeed excited in \textsc{zeus-mp} simulations,
substantiating the claim that global MHD mountain oscillations are
Alfv\'{e}nic \citep{Payne07}.
For $M_a=1.4 M_c$ (dotted curve in
\fref{spectrum:spectrum_comparison}), we see a continuous spectrum at
$f\ga 7$ Hz, albeit less distinctly than in the $M_a=M_c$ case. The $M_a=0.6 M_c$ simulation (dashed)
shows a peak at $f=23$ Hz and no obvious continuous spectrum.

\change{We do not attempt a unambiguous identification of the
  oscillation modes seen in the above simulation. As an integrated
  quantity, the behaviour of $\epsilon$ is determined by a global
  superposition of different eigenmodes, both, continuous and
  discrete. These eigenmodes are stochastically excited through
  numerical inaccuracies \citep{Vigelius08} in the simulation displayed in 
  \fref{spectrum:spectrum_comparison} and identifying the
  frequency spectrum of $\epsilon$ with the underlying eigenfunctions
  is hence difficult. The calculations undertaken in this article are
  but a   first step towards a thorough investigation of the complete
  eigenvalue problem.}

In the context of future gravitational wave observations of magnetic
mountains in low-mass X-ray binaries \citep{Melatos05}, the MHD oscillation spectrum
enters the gravitational wave signal
through sidebands and broadening near the Fourier
peaks at $f_\ast$ and $2 f_\ast$, where $f_\ast$ is the spin frequency
of the neutron star. \citet{Vigelius08b} showed  that the sidebands
can be observed in principle with next-generation
interferometers. However, in order to exploit these features fully to
probe the physics of surface magnetic fields on accreting neutron
stars, we need to know more about how the oscillations are excited
(e.g. by the variable accretion torque) and damped. The analytic
technique in this article is a useful tool for such investigations.

\change{The oscillation modes may be perpetually re-excited,
  e.g. through starquakes, variable accretion
  torques \citep{Lai99}, and possibly cyclonic flows during type I
  X-ray bursts \citep{Spitkovsky02}. \citet{Vigelius08b} show that a
  perturbation of the fluid density causes a fractional change in the
  signal-to-noise ratio $d$ of $\delta 
  d/d \approx n/N$. Similarly,  in a (very crude) 
  model where the pulse shape is determined by the positions of the 
  footpoints of the magnetic field lines at the polar cap boundary, the
  fractional change in pulse parameters (e.g. full-width
  half-maximum) is comparable to the fractional perturbation
  amplitude. Unfortunately, the excitation mechanisms are 
  poorly understood and it is unclear if the amplitude of the
  perturbation is sufficient to cause observable features in the
  gravitational wave spectrum. Ultimately, gravitational wave
  observations will yield valuable information about the underlying
  excitation physics.}

\citet{Glampedakis07} investigated ideal-MHD modes in magnetars,
taking into account the coupling between the fluid core and the
elastic crust. They found that global core-crust modes can explain
quasi-periodic oscillations (QPOs) observed during giant flares in
the soft gamma-ray repeaters SGR 1806$-$20 and SGR 1900$+$14. In contrast,
\citet{Levin06} argued that continuous coupled modes in magnetars decay
too rapidly to account for the observed QPOs. This debate was reviewed
recently by \citet{Watts07}. Using a series expansion,
\citet{Lee07,Lee08} calculated the discrete eigenmodes of
magnetars. While low frequency QPOs can be identified with fundamental
toroidal torsional modes, higher frequency ($100 \la f \la 1000$ Hz)
can be attributed to a variety of modes, such as spheroidal shear
modes or core/crust interfacial modes.

The aim of this paper is to interprete analytically and physically the
magnetic mountain oscillations seen in nonrotating numerical simulations
\citep{Payne07, Vigelius08b}. For rapidly rotating objects with
$\Omega/2 \pi \ga 0.1$ kHz, like accreting millisecond pulsars, the
continuous spectrum is strongly modified by rotation, as shown in
\sref{spectrum:splitting}. We will calculate the rotational splitting
in a forthcoming paper.

\bibliography{analytic.bib}

\end{document}